\documentclass[journal=nalefd,manuscript=letter]{achemso}
\usepackage[version=3]{mhchem}
\usepackage{amsmath}
\usepackage{amsfonts}
\usepackage{amssymb}
\usepackage{graphicx}
\usepackage{color}

\title[] {Regaining a spatial dimension: Mechanically transferrable two-dimensional InAs nanofins grown by selective area epitaxy}

\author{J.~Seidl}
\affiliation{School of Physics, University of New South Wales, Sydney NSW 2052, Australia}

\author{J.G.~Gluschke}
\affiliation{School of Physics, University of New South Wales, Sydney NSW 2052, Australia}

\author{X.~Yuan}
\affiliation{Department of Electronic Materials Engineering, Research School of Physics and Engineering, The Australian National University, Canberra ACT 2601, Australia}
\alsoaffiliation[Current address: ]{Hunan Key Laboratory for Supermicrostructure and Ultrafast Process, School of Physics and Electronics, Central South University, 932 South Lushan Road, Changsha, Hunan 410083, P.~R.~China}

\author{S.~Naureen}
\affiliation{Department of Electronic Materials Engineering, Research School of Physics and Engineering, The Australian National University, Canberra ACT 2601, Australia}
\alsoaffiliation[Current address: ]{IRnova AB, Electrum 236, Kista SE-164 40, Sweden}

\author{N.~Shahid}
\affiliation{Department of Electronic Materials Engineering, Research School of Physics and Engineering, The Australian National University, Canberra ACT 2601, Australia}
\alsoaffiliation[Current address: ]{Finisar Sweden AB, Bruttov\"{a}gen 7, J\"{a}rf\"{a}lla SE-175 43, Sweden}

\author{H.H.~Tan}
\affiliation{Department of Electronic Materials Engineering, Research School of Physics and Engineering, The Australian National University, Canberra ACT 2601, Australia}

\author{C.~Jagadish}
\affiliation{Department of Electronic Materials Engineering, Research School of Physics and Engineering, The Australian National University, Canberra ACT 2601, Australia}

\author{A.P.~Micolich}
\email{adam.micolich@nanoelectronics.physics.unsw.edu.au}
\affiliation{School of Physics, University of New South Wales, Sydney NSW 2052, Australia}

\author{P.~Caroff}
\affiliation{Department of Electronic Materials Engineering, Research School of Physics and Engineering, The Australian National University, Canberra ACT 2601, Australia}
\alsoaffiliation[Current address: ]{Microsoft Quantum Lab Delft, Delft University of Technology, 2600 GA Delft, The Netherlands}

\date{\today}

\begin{document}

\begin{abstract}

We report a method for growing rectangular InAs nanofins with deterministic length, width and height by dielectric-templated selective-area epitaxy. These freestanding nanofins can be transferred to lay flat on a separate substrate for device fabrication. A key goal was to regain a spatial dimension for device design compared to nanowires, whilst retaining the benefits of bottom-up epitaxial growth. The transferred nanofins were made into devices featuring multiple contacts for Hall effect and four-terminal resistance studies, as well as a global back-gate and nanoscale local top-gates for density control. Hall studies give a 3D electron density $2.5~-~5 \times 10^{17}$~cm$^{-3}$, corresponding to an approximate surface accumulation layer density $3~-~6 \times 10^{12}$~cm$^{-2}$ that agrees well with previous studies of InAs nanowires. We obtain Hall mobilities as high as $1200$~cm$^{2}$/Vs, field-effect mobilities as high as $4400$~cm$^{2}$/Vs and clear quantum interference structure at temperatures as high as $20$~K. Our devices show excellent prospects for fabrication into more complicated devices featuring multiple ohmic contacts, local gates and possibly other functional elements, e.g., patterned superconductor contacts, that may make them attractive options for future quantum information applications.

{\bf Keywords:} Nanofin, Selective area epitaxy, Nanowires, Hall effect
\end{abstract}

\maketitle

Quantum devices were underpinned for several decades by the interfacial two-dimensional (2D) electron gas found in III-V semiconductor heterostructures.\cite{FerryBook09} A top-down approach to these systems is costly, with heterostructure complexity limited by interfacial strain issues. Bottom-up approaches have thus generated massive interest with a heavy focus on one-dimensional (1D) nanostructures, i.e., nanowires, where small interfaces enable greater heterostructure versatility, including the ability to integrate III-Vs on low-cost Si substrates.\cite{HuACR99, SamuelsonMT03, RielMRSBull14} Researcher ingenuity has meant clever new devices still arise from the nanowire geometry even after two decades. That said, we suspect we are not alone in wishing for extra spatial dimensions to work with. An attractive idea would be to take the hexagonal nanowire cross-section and stretch it to obtain a 2D `nanofin' such that two side-facets have much larger area. These could be transferred to a separate substrate to make devices featuring, e.g., multiple contacts and gates by conventional nanofabrication methods. This concept is impossible with vapor-liquid-solid approaches.\cite{WagnerAPL64, HirumaJAP95} Here we demonstrate it is possible using selective-area epitaxy,\cite{TomiokaJMR11, GuniatChemRev19} giving 2D InAs nanofins with precise size control, and opening a path to more interesting nanostructure shapes via appropriate mask design.\cite{WangMIP19}

Our 2D nanofins offer some interesting potential for nanoelectronics. Firstly, they offer a new route to complex material geometries, e.g., the hash-tag structures recently developed towards topological braiding of Majorana zero modes,\cite{GazibegovicNat17} via established methods such as etching rather than exotic growth strategies. Secondly, the additional dimension means nanofins are better suited to making quantum devices featuring multiple contacts for Hall and/or four-terminal measurements and multiple gates for separating conduction channels or device regions. Improved contact arrangements facilitate better understanding of materials by enabling us to measure transport mobility versus carrier density rather than resort to single-figure metrics, e.g., field-effect mobility, that are used by necessity in nanowires due to contact limitations.\cite{BlomersAPL12} Finally, by depositing patterned superconductor films and exploiting electron density accumulation at the facet corners\cite{HeedtNanoscale15, DegtyarevSciRep17} at opposite edges of the nanofin, exciting new pathways to Majorana/parafermion zero-mode devices\cite{StanescuJPCM13, AliceaARCMP16} for topological quantum computation applications may be possible.\cite{NayakRMP08}

III-V nanowires were originally and are still commonly grown from a nanoparticle catalyst using a vapor-liquid-solid (VLS) approach.\cite{WagnerAPL64, HirumaJAP95} More recently, self-catalysed VLS growth has been developed.\cite{MandlNL06, FontcubertaAPL08} We use an alternative approach called selective-area epitaxy (SAE) that involves using a patterned amorphous dielectric layer to template growth on a crystalline substrate.\cite{TomiokaJMR11, GuniatChemRev19} This method, first developed for growing Si-on-Si\cite{JoyceNat62} and quickly extended to GaAs\cite{TauschJECS65} in the 1960s, was only widely used for III-V nanowires after work by Poole {\it et al.}\cite{PooleAPL03} and Motohisa {\it et al.}\cite{MotohisaJCG04} in 2003/2004. VLS growth remains popular due to historical momentum and because it provides the only route to stacking-fault-free nanowires.\cite{ChiNL13} However, the VLS method is limiting in the quest to extend beyond 1D structures. Under appropriate VLS growth conditions `sail-like' two-dimensional (2D) structures will grow as extensions from a 1D nanowire `mast'.\cite{AagesenNatNano07, delaMataNL16, PanNL16, KelrichNL16, PanNL19, KangNL19} In each case these structures have significant non-uniformity in shape, dimensions or orientation across a single growth. They also come with a nanowire `stem' and/or catalyst particle attached; the elimination of either or both would be desirable from a utopian device design perspective.

Selective-area epitaxy offers a more promising path to functional `bottom-up' 2D structures for electronic devices, giving precise and reliable deterministic control over shape, thickness and crystal structure without the baggage of catalyst particles and nanowire stems. Conesa-Boj {\it et al.}\cite{Conesa-BojACSNano12} obtained V-shaped nanomembranes by molecular beam epitaxy using nanoscale apertures in a SiO$_{x}$ mask. More recently, trench structures in a SiO$_{x}$ mask have been used to grow long horizontally-oriented InAs,\cite{HsuNRL12, GoothNL17, KrizekPRM18, VaitiekenasPRL18, LeeArXiv18, AseevNL19} GaN,\cite{YehAPL12} GaAs,\cite{ChiNL13, TutuncuogluNanoscale15, AseevNL19} and InSb\cite{DesplanqueNanotech18} nanowires, along with more exotic materials.\cite{delaMataNL19} These structures remain on their growth substrate for use as photonic structures\cite{YehAPL12, ChiNL13, TutuncuogluNanoscale15, YangNL18, StutzPSSRRL19, GuniatACSNano19}, electronic wires\cite{GoothNL17, KrizekPRM18, VaitiekenasPRL18, DesplanqueNanotech18} or as templates for further growth, e.g., InAs nanowires atop GaAs nanomembranes.\cite{FriedlNL18}

Our focus sits in a currently untapped space between the works described above -- we seek the large open areas of the 2D `sail-like' structures from catalyst-driven VLS growth but with the precise shape control and uniformity available from selective-area epitaxy and the ability to transfer the structures to a separate substrate for device fabrication. Here we report the growth and characterisation of tall, long and thin 2D InAs nanofin structures, like those in Fig.~1a-c, using dielectric-templated selective-area epitaxy. Our method produces rectangular nanofins with precise control over all three geometric dimensions. These nanofins can be mechanically-transferred to a separate substrate for fabrication into devices featuring multiple contacts and electrostatic gate structures. The geometry readily enables characterisation via Hall effect and devices with four-terminal contact arrangements for contact-resistance-corrected measurement. Our nanofins give electron transport mobilities up to $1200$~cm$^2$/Vs at typical 3D electron density $2.5~-~5 \times 10^{17}$~cm$^{-3}$ at temperature $T = 300$~mK, tunable electron density via electrostatic gating and clear quantum interference structure for $T < 20$~K. Our work opens a path to a range of more versatile and complex quantum device structures using the `bottom-up' approach.

\begin{figure}
\includegraphics[width=16cm]{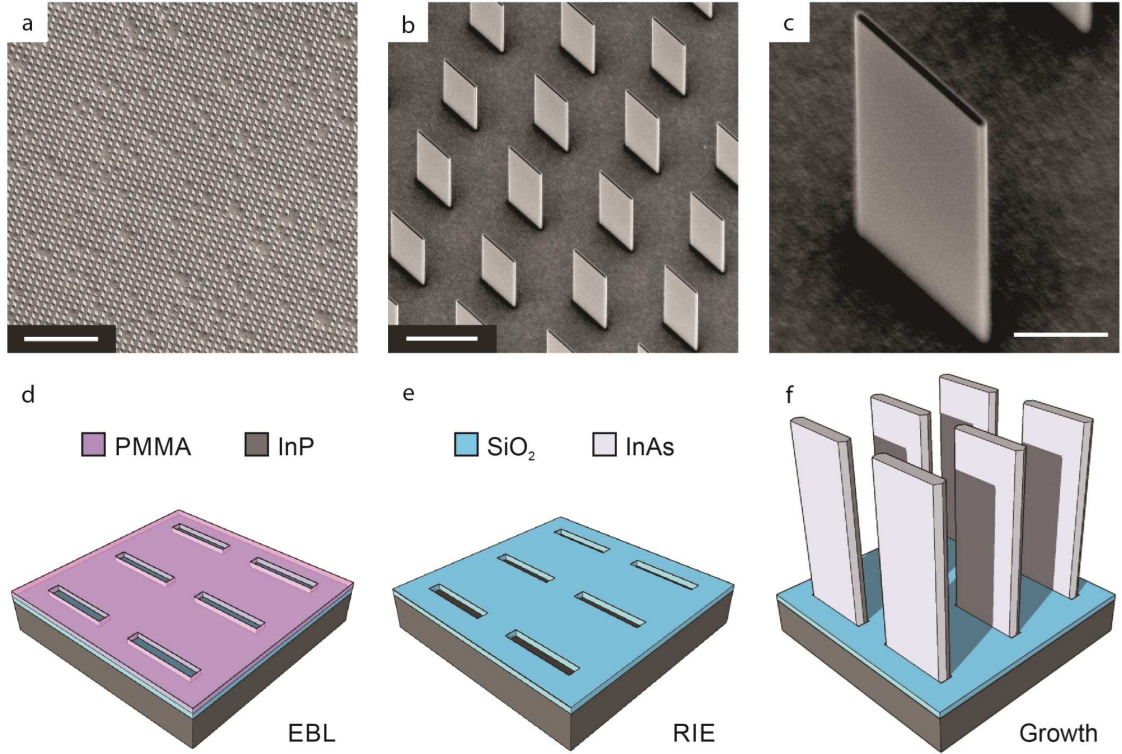}
\vspace{5mm}
\caption{{\bf Templated growth of 2D nanofin structures} {\bf a - c} Scanning electron micrographs of 2D nanofins post-growth and prior to transfer to a device substrate: {\bf a} wide-frame showing a large array of identical rectangular structures, {\bf b} zoom-in of the field in {\bf a} showing finer detail, and {\bf c} zoom-in on a single nanofin to highlight the hexagonal structure featuring two large $\{110\}$ and four smaller $\{110\}$ facets on the sides and a $\{111\}$B facet at the top. The scale bars for {\bf a}, {\bf b} and {\bf c} represent $20~\mu$m, $1.5~\mu$m and $500$~nm respectively. All images at $30^{\circ}$ tilt from perpendicular to substrate. {\bf d-f} Schematic of key steps in the template fabrication and growth process, which involves starting with a SiO$_{x}$-coated InP$(111)$B substrate (blue on dark grey), spin-coating a PMMA resist (pink), {\bf d} defining the template openings by electron-beam lithography, {\bf e} a \ce{CHF3} reactive ion etch to transfer the pattern to the SiO$_{x}$ layer followed by resist removal, and finally {\bf f} growth of InAs (light grey) by metal-organic vapor phase epitaxy. More complete details are given in Methods.}
\end{figure}

{\bf Templated growth of 2D rectangular InAs nanofins} Figures~1a-c show scanning electron micrographs of typical growth results. Figure~1a demonstrates 2D structures can be grown in large arrays with high yield ($>80\%$) and good shape uniformity. Figures~1b/c show sequential zoom-ins of the nanofins, which have typical length $\sim 1~\mu$m, width $\sim 80$~nm and height $\sim 4~\mu$m (see Supplementary Fig.~S1a/b). The structure is essentially a nanowire stretched along one symmetry axis, featuring two large $\{110\}$ face-facets and four smaller $\{110\}$ edge-facets (see Fig.~1c and Fig.~S1b). The top-facet is $\{111\}$B matching the substrate. The structure maintains the shape imposed by the mask during growth for reasons similar to those governing SAE growth of nanowires;\cite{MotohisaJCG04} the $\{111\}$B surface has a high growth rate while the $\{110\}$ surfaces provide poor nucleation suppressing lateral growth.\cite{AlbaniPRM18} Figures~1d-f highlight key steps in the template fabrication and growth process, which begins with a InP$(111)$B wafer (dark grey). This substrate was cleaned and $25$~nm SiO$_{x}$ (blue) was deposited by plasma-enhanced chemical vapor deposition. Dielectric-template patterning was performed via a mask transfer process using poly-methylmethacrylate (PMMA) electron-beam lithography (EBL) resist (pink). The mask pattern was written with a $20$~kV electron beam using a Raith-$150$ EBL system and developed in $1:3$ methylisobutylketone:2-propanol to expose the SiO$_{x}$ surface in regions where growth should occur (Fig.~1d). This pattern was then transferred to the SiO$_{x}$ by \ce{CHF3} reactive-ion etching (RIE) to reveal the InP surface at locations where the SiO$_{x}$ was exposed. The PMMA was then removed leaving the patterned SiO$_{x}$ template (Fig.~1e). All template holes have their long axis aligned with the InP$(111)$B substrate $\langle 112 \rangle$ direction unless otherwise specified. The final stage was growth of InAs (light grey) by metal-organic vapor phase epitaxy (MOVPE), with nucleation and epitaxial growth occurring at the exposed InP surfaces, giving structures shaped by the SiO$_{x}$ template (Fig.~1f). Further process details are given in the Methods section.

\begin{figure}
\includegraphics[width=16cm]{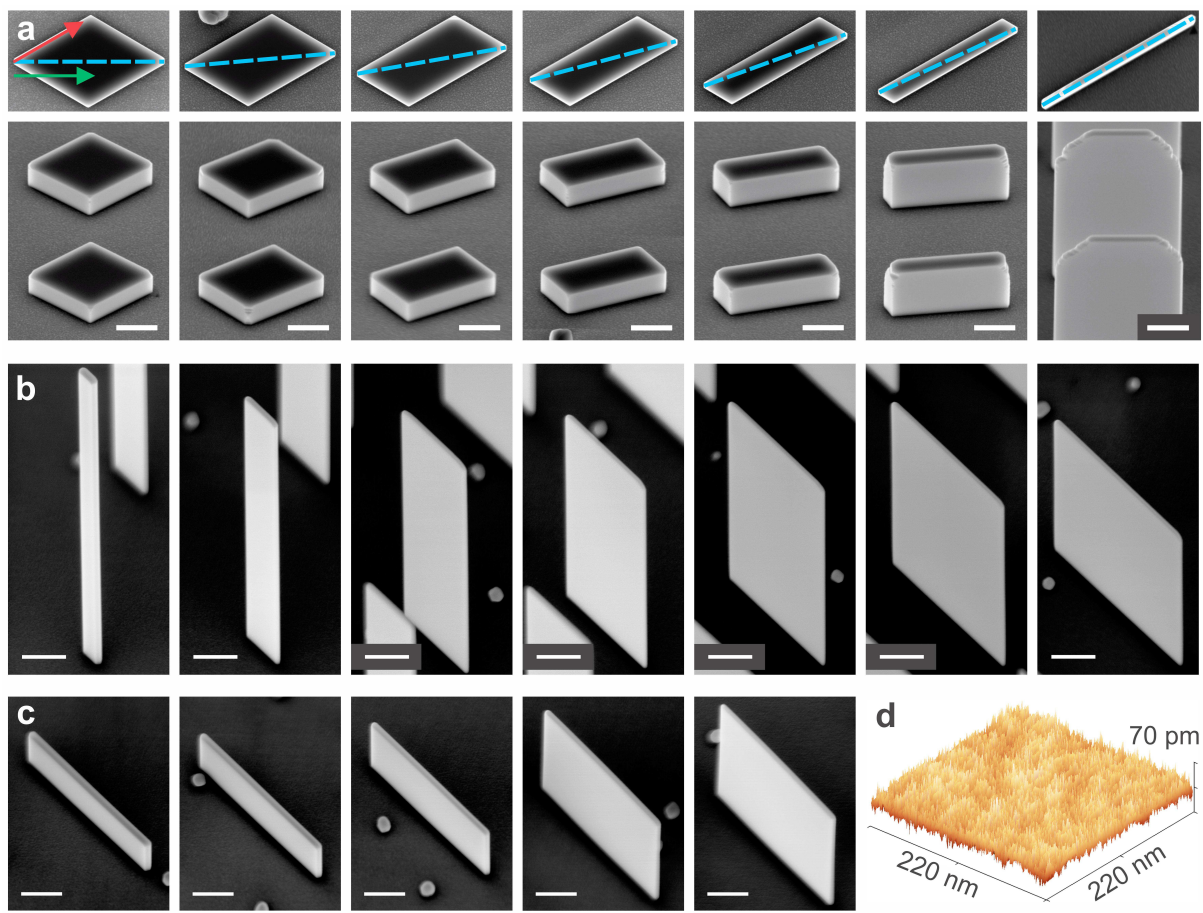}
\vspace{5mm}
\caption{{\bf Exerting control over structure via template structure} {\bf a} Overhead (top) and angled (bottom) SEM images of the growth outcome for a sequence of rectangular openings (blue dashed line) at $0^{\circ}$, $5^{\circ}$, $10^{\circ}$, $15^{\circ}$, $20^{\circ}$, $25^{\circ}$ and $30^{\circ}$ relative to the $\langle 110 \rangle$ substrate direction (green arrow). The $\langle 112 \rangle$ direction (red arrow) is shown for reference. {\bf b/c} Angled SEM images of growth outcome for {\bf b} different opening lengths $300$~nm (left), $500$~nm, $800$~nm, $1~\mu$m, $1.3~\mu$m, $1.5~\mu$m and $1.8~\mu$m (right) and {\bf c} different opening widths $120$~nm (left), $110$~nm, $100$~nm, $90$~nm and $80$~nm (right). All scale bars in {\bf a, b} and {\bf c} represent $500$~nm. {\bf d} AFM image of a nanofin surface demonstrating the flatness of the large $\{110\}$ facets. The RMS surface roughness is $80$~pm.}
\end{figure}

Figures~2a-c demonstrate three aspects of the template that affect the structures grown. Firstly, the shape is reliant on the rectangular opening's long-axis orientation relative to the underlying InP$(111)$B substrate's crystallographic axes. The two key surface directions in Fig.~2a are $\langle 110 \rangle$ (green arrow) and $\langle 112 \rangle$ (red arrow). The mask opening orientation is indicated by the blue dashed line in Fig.~2a, and is rotated in $5^{\circ}$ steps from $\langle 110 \rangle$ (far left) to $\langle 112 \rangle$ (far right). All structures grown have six $\{110\}$ side-facets and a $\{111\}$B top-facet demonstrating a strong preference to $\{110\}$ facet formation, as found for SAE-grown InAs nanowires.\cite{MotohisaJCG04, MandlNL10} For the $\langle 110 \rangle$-aligned opening, two of the $\{110\}$ facets are very small whilst the remaining four have equal size, giving a rhomboid appearance. As the opening is rotated, two of the four large $\{110\}$ facets grow while the other two shrink. Once the opening aligns with $\langle 112 \rangle$ the structure consists of two large face-facets and four small edge-facets with equal size, giving the 2D nanofins we focus on for the remainder of this work. Figures~2b and 2c show the effect of changing opening length $l$ (long axis) and width $w$ (short axis) for $\langle 112 \rangle$-aligned openings. The series in Fig.~2b clearly demonstrates nanofins are a natural evolution of nanowires, which would be obtained for $l = w$,\cite{MotohisaJCG04, MandlNL10} into the regime where $l >> w$. Figure~2c points to our tall freestanding nanofins being an extension of the horizontal SAE-grown nanowires~\cite{GoothNL17, KrizekPRM18} taken into the limit of small $w$ and long growth time. The small $w$ involved makes our 2D nanofins challenging to grow -- proper nucleation and growth require the opening floor to be very clean and $w$ needs to be constant along the opening length, both become tougher prospects as $w$ is reduced. Examples of growth when the mask is not well optimized are shown in Supplementary Fig.~S2. Even when satisfactory growth occurs, mask opening width variations at the few-nm level can significantly affect aspect ratio and surface area, dominating over more typical control parameters, e.g., temperature and V/III ratio. This occurs because this approach requires mask opening widths ($20-30$~nm) at the limit of conventional EBL, and the growth physics for free-standing III-V nanofins/membranes is complex and currently only well characterized for GaAs.~\cite{AlbaniPRM18} The observed variability in an array of nominally identical openings is addressed in Supplementary Fig.~S3.

The nanofin height decreases as the opening's long-axis is rotated away from $\langle 112 \rangle$ or $l$ or $w$ is increased, consistent with surface-diffusion controlled growth. Dimensions for the images in Fig.~2a-c are given in Supplementary Fig.~S4. Predicting the final grown height is challenging because one also needs to consider the mask opening spacing and growth conditions, e.g., temperature and V/III ratio. The spacing dependence is itself non-trivial compared to, e.g., honeycomb arrays of nanowires with hexagonal cross-section, where the spacing is single-valued. Here structural and array symmetries are both broken meaning at least four parameters are required: width, length, and separations in the width and length directions. We can however make some general observations. Firstly, comparable capture area leads to comparable added volume with larger mask opening area giving reduced height under fixed growth conditions and time. The relationship is slightly non-linear though because of adatom capture onto the growing structure occurring in addition to adatom capture onto the dielectric mask. Secondly, since these structures are strongly affected by surface diffusion on the mask and in the openings, placing nanofins in close proximity will eventually reduce the axial growth rate due to competition for In adatoms.

Turning to structural aspects, the nanofin oriented along $\langle112\rangle$ (rightmost in Fig.~2a) shows a highly stepped top-facet unlike other nanofins in Figs.~1 and 2b/c, and was grown at lower temperature and V/III ratio. The stepped top-facet arises from a kinetic limitation to the axial growth rate that depends on both the top surface area and the growth conditions, as evident in Supplementary Fig.~S5. The likelihood of top-facet stepping increases with $\{111\}$ top-facet surface area under fixed growth conditions. At fixed top-facet surface area, the incidence of top-facet stepping decreases for conditions favouring enhanced axial growth rate, namely higher temperature and higher V/III ratio. The 2D nanofins show wurtzite-zincblende polytypism, as found for InAs nanowires (for HRTEM data see Supplementary Fig.~S6).\cite{CaroffNatNano09} Nonetheless, the large $\{110\}$ side-facets have high flatness, as shown previously on SAE-grown InAs nanowires by STM.\cite{HjortNanoscale15} Figure~2d shows an AFM micrograph of the large $\{110\}$ side-facet, the RMS surface roughness is $\sim 80$~pm compared to $295$~pm for the underlying \ce{SiO2} device substrate surface.

\begin{figure}
\includegraphics[width=16cm]{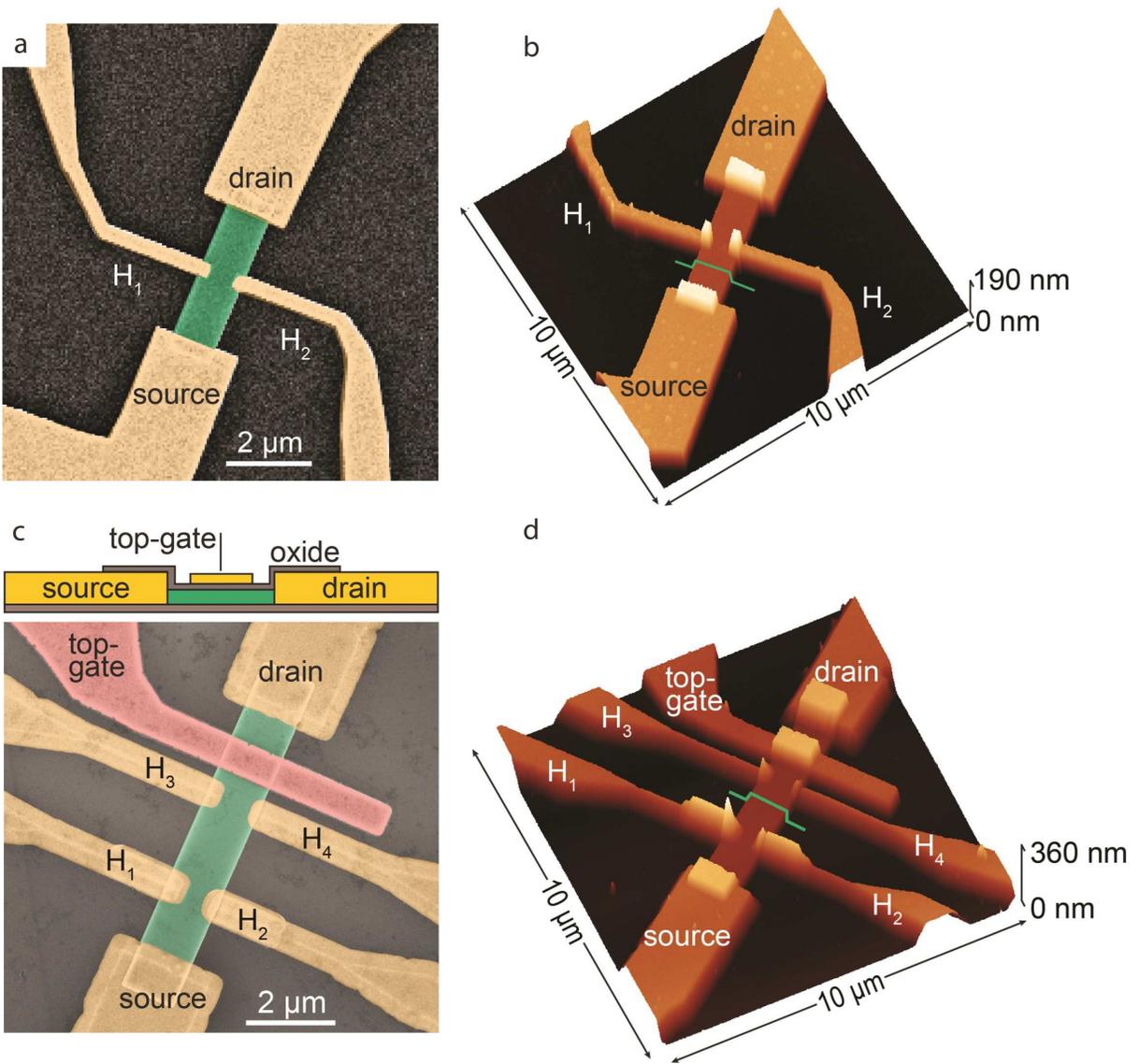}
\vspace{5mm}
\caption{{\bf Fabrication of nanofin devices for Hall effect and local gating studies} {\bf a} false-color SEM and {\bf b} AFM image of a device for Hall effect studies (Device~1) featuring a nanofin (green), source, drain and a pair of Hall contacts H$_1$ and H$_2$ (yellow). {\bf c} Schematic and false-color SEM and {\bf d} AFM image of a patterned top-gate device (Device~2) featuring nanofin (green), set of six contacts contacts (yellow), and a \ce{HfO2} insulated top-gate (red). The scale bars in {\bf a} and {\bf c} represent $2~\mu$m. The green dotted lines in {\bf b} and {\bf d} indicate the locations of an AFM line-scan revealing nanofin thicknesses of $74$~nm and $85$~nm for the two devices, made from separate growths (see Supplementary Fig.~S7).}
\end{figure}

{\bf Mechanical transfer of nanofins and device fabrication} A key motivation was to obtain 2D structures for transfer to a separate substrate for fabrication into devices with multiple gates and contacts. We previously used dry-transfer via a small triangle of lab-wipe for nanowire devices.\cite{StormNL12} This works acceptably but is brutal and costly -- the large tip rapidly decimates a field like that in Fig.~1a, which requires a very long EBL session for writing the growth template. Nanoimprint lithography might help alleviate this cost issue.\cite{MunshiNL14} Wet deposition involving ultrasonication into solvent is also expensive because large arrays are needed to obtain feasible liquid volume with suitable nanofin concentration. Instead we perform deposition using a micromanipulator mounted on a high-resolution optical microscope.\cite{GazibegovicNat17, delaMataNL16} This enables transfer of single nanofins with a positional accuracy of order $10~\mu$m, high yield ($\sim 80\%$) and minimal growth field decimation. The ease of detaching a nanofin improves with increased height and/or decreased base length. With care, good technique and patience, nanofins can mostly be cleaved cleanly at the base, enabling the entire nanofin to be transferred.

Device fabrication thereafter proceeds by conventional methods. The device substrate was a $n^{+}$-Si wafer with a $100/10$~nm thick \ce{SiO2}/\ce{HfO2} insulator and pre-patterned Ti/Au interconnect and alignment structures. The $n^{+}$-Si substrate was used as a back-gate for all devices. The substrate was cleaned and nanofins were transferred mechanically using a micromanipulator to give a few transferred nanofins per $100 \times 100~\mu$m active device field on the substrate. The transferred nanofins adhere strongly by van der Waals forces. We spin-coat PMMA resist prior to defining source and drain leads and Hall probes using EBL. Contacts were passivated with \ce{(NH4)2S}$_{x}$ solution prior to thermal evaporation of approximately $10/150$~nm Ni/Au and lift-off to give the completed device in Fig.~3a,b (Device~1). Top-gate structures can be added thereafter. This was achieved with a further two rounds of EBL. First we pattern a gate-insulator, which is approximately $12-20$~nm of \ce{HfO2} or \ce{Al2O3} by atomic-layer deposition (ALD), followed by lift-off. Then we pattern gates, which are approximately $10/135$~nm Ti/Au by vacuum thermal evaporation, followed by lift-off. This gives the completed device in Fig.~3c,d (Device~2). Full details are in the Methods section with specific values for each device tabulated in Supplementary Table~1.

\begin{figure}
\includegraphics[width=16cm]{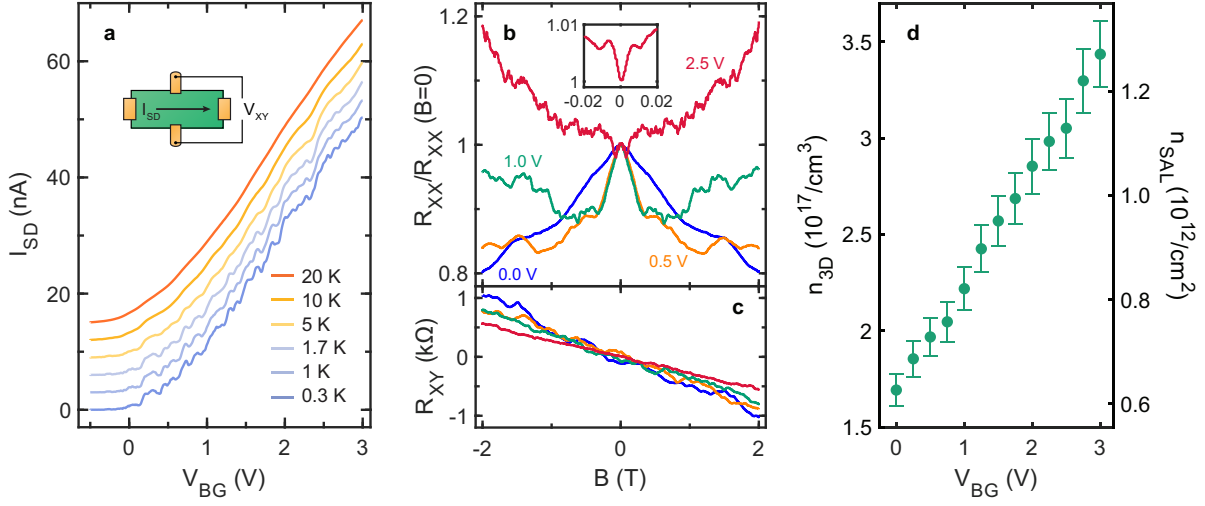}
\vspace{5mm}
\caption{{\bf Electrical characterisation of nanofin Hall device} {\bf a} Source-drain current $I_{SD}$ vs back-gate voltage $V_{BG}$ as a function of temperature $T$ obtained for the `normal' configuration. The `rotated' configuration data appears in Supplementary Fig.~S8/9 as discussed in the text. Consecutive traces are offset upwards by $3$~nA as $T$ is increased for clarity (lowest $T$ trace has zero offset). {\bf b} zero-field-scaled longitudinal resistance $R_{XX}/R_{XX}(B = 0)$ and {\bf c} Hall resistance $R_{XY}$ vs magnetic field $B$ at four different back-gate voltages $V_{bg} = 0$~V (blue), $+0.5$~V (orange), $+1$~V (green) and $+2.5$~V (red) obtained at $T = 300$~mK. Inset: Focus on low $B$ for the $V_{BG} = +2.5$~V data to highlight the $R_{XX}/R_{XX}(B = 0)$ minima at $B = 0$ (see Supplementary Fig.~S10 for all traces over this $B$-range). {\bf d} Measured 3D electron density $n_{3D}$ (left axis) and corresponding approximate surface accumulation layer density $n_{SAL}$ (right axis) vs $V_{BG}$ obtained from Hall effect data. All data were obtained from Device~1 in the normal orientation.}
\end{figure}

{\bf Electrical characterisation of Hall-configuration nanofin device} We began by studying nanofin structures featuring a pair of Hall contacts and a back-gate (Device~1) as shown in Fig.~3a/b. The nanofin forms a channel $1~\mu$m wide and $3.5~\mu$m long with the two Hall probes on opposing sides approximately half-way along the nanofin. Electrical measurements were performed in an Oxford Instruments Heliox VT $^3$He cryostat with $2$~Tesla superconducting magnet using standard a.c. lock-in techniques. Before discussing the data, we preemptively highlight some aspects of our conduction channel that are important to understanding these devices. A well-known feature of InAs is the tendency for surface states to pin the surface Fermi energy at the conduction band edge, giving a surface accumulation layer (SAL) with high electron density.\cite{NoguchiPRL91} Electronic structure calculations for nanowires point to the SALs for the six $\{110\}$ facets joining to form a hexagonal-cylinder geometry, with slightly higher electron density at the corners between adjacent facets.\cite{HeedtNanoscale15, DegtyarevSciRep17} However, several experiments indicate conduction is not solely via this SAL, with significant transport via the nanowire core,\cite{SchefflerJAP09, BlomersNL11, JespersenPRB15} where free carrier density is likely only an order of magnitude smaller at most.~\cite{HeedtNanoscale15} Thus a sensible expectation is for an inhomogeneous 3D electron distribution featuring slightly higher density SALs, potentially with poor mobility due to surface proximity, and a lower density core with higher mobility due to screening by the SALs. This explains why, in what follows, we {\it a priori} treat our measurements from a 3D perspective.

Figure~4a shows the source-drain current $I_{SD}$ in response to source-drain voltage $V_{SD} = 500~\mu$V versus back-gate voltage $V_{BG}$ between $T = 280$~mK and $20$~K. For completeness we obtained data for both possible Hall configurations. Data for the `normal' orientation is presented in Fig.~4. Data for the `rotated' orientation where $V_{SD}$ is applied and $I_{SD}$ passed via H$_1$ and H$_2$ is presented in Supplementary Figs~S8/9 to provide additional insight into the transport. Starting with Fig.~4a, negative/positive $V_{BG}$ leads to reduced/increased $I_{SD}$ (depletion/enhancement) consistent with electrons as the majority carrier. The device has relatively low conductivity at $V_{BG} = 0$~V but this is not unexpected at low temperatures. A positive shift in gate threshold upon cooling, and ultimately a positive threshold voltage at low temperature, is commonly seem in past studies of InAs nanowires.\cite{TianNL12, WuNL13, GoothNL17} The low temperature data in Fig.~4a shows reproducible quantum interference fluctuations that reduce in amplitude with increasing temperature, consistent with observations in both 1D InAs nanowires\cite{ThelanderSSC04, LiangNL09} and 2D open quantum dots in GaAs.\cite{BirdPRL99} The fluctuations remain visible up to $T \sim 10$~K, indicating long electron phase coherence length, and are stronger for the rotated orientation due to the reduced contact separation (see Fig.~S8).

Figures~4b/c show the zero-field-scaled longitudinal magnetoresistance $R_{XX}/R_{XX} (B = 0)$ where $R_{XX} = V_{SD}/I_{SD}$ and Hall resistance $R_{XY} = V_{H}/I_{SD}$ versus magnetic field $B$ for four different $V_{BG}$ values. Corresponding data for the rotated orientation appears in Supplementary Fig.~S9. The $R_{XX}/R_{XX} (B = 0)$ traces show structure reminiscent of open quantum dots,\cite{BirdRPP03} with a central magnetoresistance peak surrounded by symmetric, reproducible quantum interference fluctuations. These fluctuations also appear in the $R_{XY}$ data. The fluctuations are suppressed with increasing $T$, and to the field range available, show no structures indicative of Shubnikov-de Haas oscillations (Fig.~4b) or quantum Hall effect (Fig.~4c), as might be expected for a large-area planar 2DEG in InAs (see also Fig.~S9).\cite{StephensPRB78} This is not surprising given the nanofin dimensions ($1~\times~3.5~\mu$m) are closer to those of an open quantum dot ($\sim 1~\times~1~\mu$m)\cite{BirdRPP03} than a conventional AlGaAs/GaAs Hall bar ($\sim 0.4~\times~1$~mm).\cite{StormerRMP99} Shubnikov-de Haas oscillations and quantum Hall plateaus were not observed in separate studies at magnetic fields up to $6$~T at $T = 4$~K either. This may simply be due to insufficient classical and quantum scattering lifetimes in our nanofins.\cite{HarrangPRB85} The lack of quantum Hall plateaus might also point to the conduction channel being insufficiently 2D\cite{vonKlitzingRMP86} due to conduction via the nanofin core. The peak at $B = 0$ in Fig.~4b is commonly observed in InAs nanowires, and often attributed to weak localization.\cite{HansenPRB05} Our peak is gradually suppressed with more positive $V_{BG}$, with $R_{XX}$ evolving a sharper central minima for $V_{BG} > +1.0$~V. The sharp central minima obtained for $V_{BG} = +2.5$~V appears inset to Fig.~4b with a more complete series in Supplementary Fig.~S10. We tentatively attribute this minima to weak anti-localization (WAL).\cite{BergmannPhysRep84} However, the superimposed quantum interference structure makes a definitive attribution of $B = 0$ maxima/minima to weak localization or antilocalization challenging, as was the case for open quantum dots.\cite{AkisPRB99, BirdRPP03} If we assume the $R_{xx}$ minima are gate-dependent WAL features, and fit using the model by Iordanskii {\it et al.}\cite{IordanskiiJETPLett94}, we obtain phase-coherence lengths $L_{\phi}$ of $200 – 600$~nm and spin-relaxation lengths $L_{SO}$ as low as $150$~nm. These values are comparable to those found for InAs nanowires.\cite{LiangNL12, HansenPRB05, TakaseSciRep17} Indicative fits and plots of $L_{\phi}$ and $L_{SO}$ versus $V_{BG}$ are presented in Supplementary Fig.~S10. A detailed study of localization/scattering in our nanofins will be the subject of a separate paper.

\begin{figure}
\includegraphics[width=16cm]{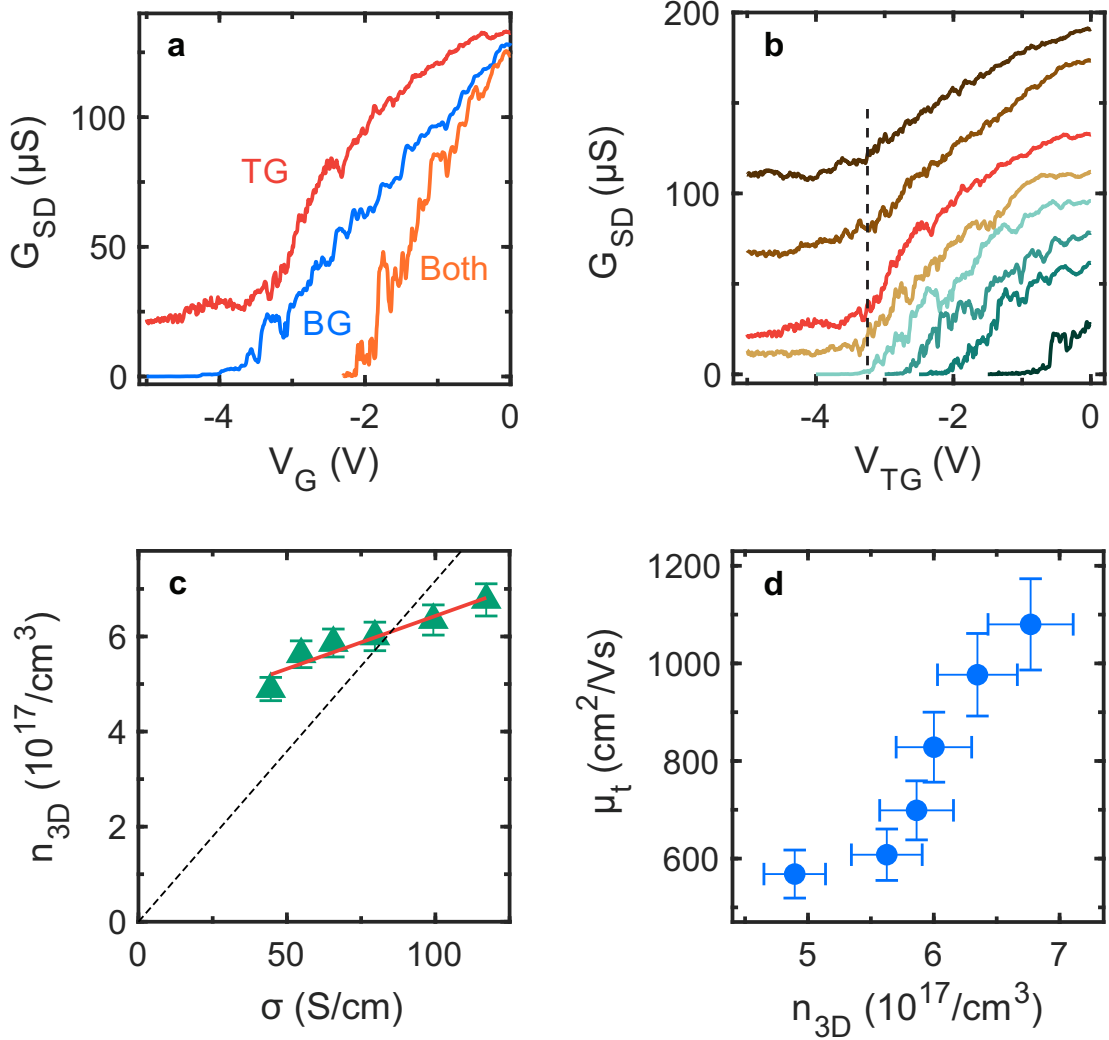}
\vspace{5mm}
\caption{{\bf Electrical characterisation of dual-gated nanofin device} {\bf a} Source-drain conductance $G_{SD} = I_{SD}/V_{SD}$ vs gate voltage $V_{G}$ for the back-gate with top-gate grounded (blue), top-gate with back-gate grounded (red), and both gates biased together (orange). {\bf b} $G_{SD}$ vs top-gate voltage $V_{TG}$ at fixed back-gate voltage $V_{BG}$ obtained at $V_{BG} = +2$~V (brown, top trace), $+1$~V, $0$~V (red), $-0.5$~V, $-1$~V, $-1.5$~V, $-2$~V and $-3$~V (green, bottom trace). The dashed line in {\bf b} is a guide to the eye. {\bf c} 3D electron density $n_{3D}$ obtained at different $V_{BG}$ vs conductivity $\sigma$ with a linear fit (orange line) and a fit forced through $\sigma = 0, n_{3D} = 0$ as per Bl\"{o}mers {\it et al}. (black dashed line)\cite{BlomersAPL12}. {\bf d} Plot of transport mobility $\mu_{t}$ vs $n_{3D}$ calculated on a single-point basis (see text). All data obtained from Device~2 at $T = 0.3$~K.}
\end{figure}

A notable aspect of our nanofins is the comparative ease in obtaining Hall measurements. This is difficult for nanowires due to the small contact gaps involved and the overhang generated by the hexagonal geometry.\cite{StormNatNano12, BlomersAPL12} Although the measurements are easier, the interpretation needs some care. As mentioned earlier, experiments point to conduction throughout the structure,\cite{SchefflerJAP09, BlomersNL11, JespersenPRB15} motivating us to start with a 3D treatment. In Fig.~4d we plot the 3D electron density $n_{3D}$ (left axis) versus $V_{BG}$ using the measured nanofin thickness $t = 74$~nm. Although conduction occurs through the entire structure, there is no avoiding that the electron density is higher closer to the nanofin surface, likely by over an order of magnitude.\cite{HeedtNanoscale15} For our geometry, the Hall voltage is dominated by two of the six side-facets. Thus we suggest the approximation $n_{SAL} \approx (n_{3D}t)/2$, where $n_{SAL}$ is the surface accumulation layer density. We provide this as the right-hand axis for Fig.~4d accordingly. However, some words of caution are warranted. The $n_{SAL}$ estimate automatically implies the top and bottom SALs have equal density, since we cannot measure them independently because they are shorted by the nanofin edge, i.e., the four SALs at the four small edge-facets. Firstly, changing $V_{BG}$ will necessarily shift charge between the top and bottom SALs so that $n_{SAL}^{top} = n_{SAL}^{bottom}$ only holds at one $V_{BG}$. Secondly, the $V_{BG}$ where $n_{SAL}^{top} = n_{SAL}^{bottom}$ can vary substantially from zero due to surface chemistry affecting surface-state density.\cite{BabadiJAP14, HeedtNanoscale15} For Device~1 both surfaces are chemically pristine, contacts aside and ignoring organic residues from lithography. But for our gated devices (Device~2), the addition of a gate-oxide by atomic layer deposition on the top nanofin surface likely means $n_{SAL}^{top}$ differs substantially from $n_{SAL}^{bottom}$ at $V_{BG} = 0$. Thus our $n_{SAL}$ is an average of the top and bottom SALs, and at best an order of magnitude estimate. Nonetheless, it is useful for comparison against earlier studies. In Fig.~4d we obtain $n_{SAL}$ values between $0.6 \times 10^{12}$~cm$^{-2}$ and $1.3 \times 10^{12}$~cm$^{-2}$ for Device~1. This agrees well with the values ranging from $0.45 \times 10^{12}$~cm$^{-2}$ to $1.3 \times 10^{12}$~cm$^{-2}$ obtained by Bl\"{o}mers {\it et al.}\cite{BlomersAPL12} for Hall measurements of InAs nanowires grown by molecular-beam epitaxy. Our values also agree to order of magnitude with capacitance-voltage measurements of InAs nanowire arrays\cite{AstromskasJAP10} and InAs wafer surfaces.\cite{NoguchiPRL91} We observe a linear decrease in $n_{3D}$ with increasingly negative $V_{BG}$ due to electron depletion.

{\bf Electrical characterisation of strip-line-gated nanofin Hall-bar device} Device~2 features six ohmic contacts in a Hall bar arrangement, a global back-gate and a $650$~nm wide \ce{HfO2}-insulated top-gate between Hall probes 3 and 4 and the drain contact, as shown in Fig.~3c/d. The contact set enables full four-terminal measurement capability for obtaining the longitudinal $R_{XX}$ and Hall $R_{XY}$ resistances independent of contact contributions.\cite{WennerBSB15, ValdesPIRE54} This is often difficult for nanowires because the contacts cross the entire conduction path causing scattering.\cite{dePicciottoNat01, ThelanderSSC04} The strip-line gate is adjacent to the drain to avoid gate metallization from affecting four-terminal transport mobility measurements; if the gate was across the middle, it would be present in the voltage path for $R_{XX}$ but not $R_{XY}$. We begin in Fig.~5a by testing independent action of the top-gate (red) and back-gate (blue). In each case the other gate is grounded. The back-gate achieves full depletion ($G_{SD} = 0$) at $V_{BG} = -4$~V whereas the top-gate only achieves partial depletion, with $G_{SD} \sim 25~\mu$S for $V_{TG}$ beyond $-3.3$~V. Notably, both traces have similar slope despite the back-gate insulator being considerably thicker ($100/10$~nm \ce{SiO2}/\ce{HfO2} for back-gate versus $12$~nm \ce{HfO2} for front-gate). If both gates are biased simultaneously (orange trace in Fig.~5a), full depletion is achieved at much lower bias, as expected.

We investigate the gating action further in Fig.~5b, where we plot $G_{SD}$ versus $V_{TG}$ at various fixed $V_{BG}$. The data from Fig.~5a at $V_{BG} = 0$~V appears in red. Corresponding data for $G_{SD}$ versus $V_{BG}$ at various fixed $V_{TG}$ is shown in Supplementary Fig.~S11. The failure of top-gating action always occurs at the same $V_{TG}$ at more positive $V_{BG}$, as highlighted by the vertical dashed line in Fig.~5b. To rule out a gate discontinuity, we put a probe needle at the far end and measured a gate strip resistance of $<~100$~ohms. Looking to the left of the dashed line in Fig.~5b, the conductance where the top-gate ceases depleting is clearly influenced by the back-gate. This indicates that the part of the conduction channel that cannot be fully depleted by the top-gate clearly can be gated from the opposite side. Interestingly, the top-gate achieves no further depletion out to $V_{TG} < -7$~V at $V_{BG} = -0.7$~V and $-0.8$~V, but achieves pinch-off at $V_{TG} \sim -4$~V at $V_{BG} = -1.0$~V (see Supplementary Fig.~S12). This suggests the loss of depletion is strong and onsets sharply. We see similar behavior, i.e., failure to achieve pinch-off in a separate device (Device~3, see Fig.~S15) with $20$~nm \ce{Al2O3} gate insulator, pointing to this being a consistent behavior in nanofin devices. One possible explanation is screening by a high free electron density in the nanofin. To examine this, we modelled our device in COMSOL Multiphysics with results presented in Supplementary Figs~S13/14. In the model we can set the free electron density $n$ throughout the nanofin at zero gate bias ($V_{BG} = V_{TG} = 0$). We present data for two densities: $n = 5 \times 10^{16}$~cm$^{-3}$ and $2 \times 10^{17}$~cm$^{-3}$ corresponding to typical measured $n_{3D}$ for our devices. At $n = 5 \times 10^{16}$~cm$^{-3}$ we see the back-gate head towards pinch-off, while the top-gate, which starts with a steeper transconductance $dI/dVg$, quickly saturates at finite $I_{SD}$ (see Fig.~S14a). This behavior exacerbates at $n = 2 \times 10^{17}$~cm$^{-3}$ with both the top- and back-gates saturating at finite $I_{SD}$ (see Fig.~S14b). We see this behavior in a separate device featuring only a pair of contacts and global top- and back-gates (Device~3). The Device~3 characteristics are shown in Supplementary Fig.~S15, where we find the top- and back-gate act weakly alone but achieve pinch-off if biased together. Comparison with the COMSOL model points to an additional aspect of Fig.~5a to explain: Why is the top-gate transconductance so poor and comparable to that of the back-gate despite the thinner high-$\kappa$ oxide? The most plausible explanation is charge trapping at the upper \ce{HfO2}/InAs interface, which is deposited by ALD, whereas the lower \ce{HfO2} interface is by van der Waals force only. The lower interface should have negligible effect on InAs surface chemistry whilst the upper interface should be radically different due to the chemistry of ALD.\cite{WheelerME09} The charge trapping effects of gate-oxides on InAs nanowires typically onset at negative gate voltage and become more pronounced with increasingly negative voltage.\cite{DayehAPL07, RoddaroAPL08, BabadiAPL16} Indeed, in Supplementary Fig.~S16 we show gate sweeps in both directions for the top-gate and back-gate on Device~2. For the back-gate we see negligible hysteresis over the entire $-4.5 < V_{BG} < 0$~V gate range. However, for the top-gate, we see the onset of hysteresis at $V_{TG} = -2$~V with it becoming very strong for $V_{TG} < -3$~V, close to where top-gate saturation occurs. This suggests charge trapping may also play a role, although our COMSOL modelling suggests we do not require trapping to explain gate saturation, which can be entirely due to screening by free electron density in the nanofin.

Together, the results above suggest the need for careful engineering of screening to implement fully operational local gates on future InAs nanofin devices. One option is to grow thinner nanofins. In our COMSOL model, effective gating can be recovered at reduced nanofin thickness $t = 40$~nm even at the higher free electron density $n = 2 \times 10^{17}$~cm$^{-3}$ (see Fig.~S14c). Another solution for thicker nanofins would be to use a global back-gate to lower the density independently of other patterned local top- or back-gates.\cite{FasthNL05, GluschkeNanotech19} Regarding the gate insulator, one possibility is to avoid ALD-deposited oxides and opt for alternative insulators, e.g., parylene.\cite{GluschkeNL18} We make one final comment regarding the data in Figs.~5a/b and Supplementary Fig.~S11. The two-stage pinch-off\cite{EisensteinAPL90} that we would expect if conduction was dominated by SALs at the top and bottom facets separated by a non-conducting nanofin core is notably absent in our device. Instead, our roughly linear gate dependencies are more consistent with a picture where conduction is more evenly spread through the nanofin with higher density but lower mobility at the surfaces and lower density with higher mobility in the core.

{\bf Four-terminal resistivity capability} We finish by using our four-terminal measurement set-up to investigate the mobility for our device. There are two possible mobilities to consider. The first is the transport mobility $\mu_{t} = \sigma/en_{3D}$, which we can obtain by using Hall measurements to get the electron density and the four-terminal resistance at $B = 0$ combined with the nanofin dimensions to get the conductivity $\sigma$. This is the mobility traditionally obtained for 2D systems. The second is the field-effect mobility $\mu_{FE} = \frac{\partial G}{\partial V_{BG}}\frac{L^{2}}{C}$, where $\frac{\partial G}{\partial V_{BG}}$ is the gate transconductance above threshold, $L$ is the channel/nanofin length and $C$ is the gate capacitance, which we obtain as $C = \frac{\epsilon_{0}LW}{d_{\ce{SiO2}}/\kappa_{\ce{SiO2}} + d_{\ce{HfO2}}/\kappa_{\ce{HfO2}}}$ with nanofin width $W$. This is the mobility more frequently used for InAs nanowires since the transport mobility cannot be readily obtained. Note also that $\mu_{FE}$ is a single value obtained in the linear region above threshold voltage whereas $\mu_{t}$ can be obtained over a wide range in gate voltage and therefore electron density.

In Fig.~5c we plot $n_{3D}$ vs $\sigma$ obtained at several different $V_{BG}$ for Device~2. A linear fit can be used to obtain $\mu_{t}$, however, in contrast to Bl\"{o}mers {\it et al.},\cite{BlomersAPL12} we find that our fit (orange line in Fig.~5c) does not pass through $n_{3d} = 0$ at $\sigma = 0$. A forced fit through $(0,0)$ is obviously poor (black dotted line in Fig.~5c). Extrapolating our unforced fit (orange line) implies that $\sigma \rightarrow 0$ at finite $n_{3D}$, an expected outcome of localization due to disorder.\cite{MottMIT74} Note also that our data is obtained at $T = 0.3$~K. This makes our thermal broadening $1000$~times smaller than for Bl\"{o}mers et al.\cite{BlomersAPL12}, where all measurements are obtained at $300$~K. Our unforced fit to the data in Fig.~5c (orange line) gives $\mu_{t} = 2800$~cm$^{2}$/Vs. This compares well to the $\mu_{t} \sim 3600$~cm$^{2}$/Vs obtained by Bl\"{o}mers {\it et al.}\cite{BlomersAPL12} for MBE-grown InAs nanowires, which should have fewer impurities than our MOVPE-grown InAs nanofins. Our $\mu_{t}$ obtained this way is likely an overestimate, it may be more correct to assume instead that $\mu_{t}$ varies with $n_{3D}$. This is not unexpected. Mobility often changes with density, for example, in an InGaAs/InAs/InGaAs heterostructure, the mobility increases with density due to screening of background impurities and native charged point defects.\cite{HatkeAPL17} Accordingly, we plot $\mu_{t}$ obtained on a single-point basis using the data in Fig.~5c, i.e., simply calculate $\mu_{t} = \sigma/en_{3D}$ for each data point, against $n_{3D}$ in Fig.~5d. The $\mu_{t}$ values range from $600-1200$~cm$^{2}$/Vs, still respectable compared to MBE-grown InAs nanowires.\cite{BlomersAPL12} We find that $\mu_{t}$ increases with $n_{3D}$, which we also attribute to screening. There are likely two contributions here: a) better screening of background impurities in the core by the higher $n_{3D}$, and b) enhanced screening of surface scattering by the SALs. A deeper study is a subject for future work, but we encourage theoretical studies of mobility versus density in these more surface-exposed structures to better understand the scattering mechanisms involved. Finally, we compare our transport mobility with field-effect mobility. For Device~2 the corresponding $\mu_{FE} = 4400$~cm$^{2}$/Vs is $2-3\times$ higher than $\mu_{t}$ (see Supplementary Fig.~S17 for underpinning data). If we compare $\mu_{t}$ with $\mu_{FE}$ for our other devices, we typically find $\mu_{FE}$ ranges from slightly above $\mu_{t}$ to several times $\mu_{t}$. Our findings are consistent with Bl\"{o}mers {\it et al.},\cite{BlomersAPL12} who also found $\mu_{FE}$ generally substantially exceeds $\mu_{t}$ due to overestimations implicit in the field-effect mobility technique.

{\bf Future prospects} Our results above demonstrate the ability to transfer nanofins to a substrate with a global back-gate, and thereafter add multiple ohmic contacts and/or patterned top-gates. There are several aspects for future work. The first is to improve the performance of patterned top-gates. This may involve reducing the nanofin thickness, engineering the gate-insulator used to reduce trapping, or perhaps replacing it entirely with an insulator that does not change the surface chemistry, e.g., parylene.\cite{GluschkeNL18} Patterned local back-gates would also be desirable. This could be achieved by positioning the nanofin over pre-patterned back-gate structures on the device substrate.\cite{FasthNL05, GluschkeNanotech19} An interesting direction is to extend beyond normal metals to superconductors towards topological quantum information applications. A current approach involves coupling a superconductor to a semiconductor nanowire with strong spin-orbit coupling, e.g., InSb,\cite{MourikSci12} InAs\cite{ChangNatNano15} or InAsSb.\cite{SestoftPRM18} More advanced designs for manipulating parafermion modes involve nanowire networks,\cite{GazibegovicNat17, KrizekPRM18} which might also be implemented by etched or gated 2D nanofin structures with patterned superconductor islands/contacts deposited on them (see, e.g., concepts in Alicea \& Fendley\cite{AliceaARCMP16}). The presence of a hard gap in the Al-on-InAs system is demonstrated,\cite{ChangNatNano15} as is the ability to achieve a hard-gap without direct epitaxial growth of superconductor-on-semiconductor.\cite{GulNL17} However, a more forward-looking option inspired by Krogstrup {\it et al.}\cite{KrogstrupNatMater15} could involve an MOVPE system load-locked to an MBE system, such that nanofins can be grown, and then transferred to high vacuum\cite{MayNJP13} without air exposure for epitaxial Al deposition onto the large nanofin facets. An additional nice aspect of the nanofins is the potential for accumulation of high electron density at the two nanofin edges because each edge has three facet corners.\cite{HeedtNanoscale15, DegtyarevSciRep17} These might provide natural 1D channels for use in parafermion-based device designs.

{\bf Conclusions} We have demonstrated a method for the growth of rectangular InAs nanofins with deterministic length, width and height by dielectric-templated selective-area epitaxy methods. These freestanding nanofins can be transferred mechanically to lay flat on a separate device substrate for fabrication into device structures. A major benefit is that we regain a spatial dimension to exploit for device design compared to nanowires, whilst retaining the benefits of the `bottom-up' epitaxial growth approach, e.g., tiny interfacial areas to enable high-quality heterostructuring.\cite{SamuelsonMT03} The transferred nanofins can be prepared into devices featuring multiple contacts for Hall effect and four-terminal resistance studies, as well as a global back-gate and nanoscale local top-gates for density control. Electrical studies of our nanofin transistors point strongly to conduction throughout the nanofin thickness, with two key contributions because the electron density is strongly inhomogeneous. Firstly, there is a high density but low mobility surface accumulation layer that facilitates ohmic contact. Conduction predominantly occurs via the nanofin core, where although the electron density is lower, the mobility should be higher due to screening of surface scattering by the surface accumulation layers. Our Hall studies reveal a 3D electron density $2.5~-~5 \times 10^{17}$~cm$^{-3}$, which corresponds to an approximate surface accumulation layer density $3~-~6 \times 10^{12}$~cm$^{-2}$, in good agreement with previous studies of InAs nanowires.\cite{BlomersAPL12, AstromskasJAP10} We obtain transport mobilities up to $1200$~cm$^{2}$/Vs and clear quantum interference structure at temperatures up to $20$~K. Our nanofins show excellent prospects for fabrication into more complicated devices featuring multiple ohmic contacts, local gates and possibly other functional elements, e.g., patterned superconductor contacts. This may make them an attractive option for future quantum information applications.
\\
\\
{\bf Methods}
\\
{\bf SAE template fabrication:} Growth was performed on undoped InP(111)B substrates. The template was $25 \pm 1$~nm of SiO$_{x}$ deposited by plasma-enhanced chemical vapor deposition (PECVD) at $300^{\circ}$C in an Oxford Plasmalab 100 system and calibrated using ellipsometry. $70$~nm EBL resist ($495$k-A2 PMMA) was spin-coated at $3000$~rpm for $60$~s and baked at $180^{\circ}$C for $3$~min on a hotplate. EBL was performed using a Raith 150 EBL system with $20$~kV beam energy and $7.5~\mu$m aperture. Development was performed in $1:3$ methylisobutylketone:2-propanol solution for $60$~s followed by $2$~min oxygen plasma ash (PVA TePla, $300$~W, $300$~sccm \ce{O2} flow) to remove any resist residue in patterned areas. Pattern transfer from the PMMA into the SiO$_{x}$ was achieved by \ce{CHF3}-based reactive ion etching in an Oxford Plasmalab 80+ system. The PMMA resist was stripped in room temperature acetone, followed by a $20$~min oxygen plasma etch (PVA TePla, $300$~W, $300$~sccm \ce{O2} flow) to ensure all organic residues were completely removed. A $5$~s dip in a $1\%$ HF solution was performed immediately prior to growth to ensure the exposed InP surfaces are oxide-free.

{\bf SAE InAs growth:} The templated substrates were transferred to an Aixtron 200/4 metal-organic vapor phase epitaxy (MOVPE) immediately after the $1\%$ HF dip noted above. A pre-growth anneal in \ce{PH3}/\ce{H2} at $750^{\circ}$C for $10$~min was performed prior to growth. Growth was performed at $550 - 725^{\circ}$C at $100$~mbar in a $14.5$~L/min \ce{H2} carrier gas flow with $35~\mu$mol/min of trimethyl indium (TMIn) for all growth runs and $0.7 - 6$~mmol/min arsine (\ce{AsH3}), giving V/III ratio between $110$ and $1000$. Growth was initiated/terminated by adding/removing the group III precursor to/from the gas flow. Cooling down to $350^{\circ}$C was done with the adequate hydride/\ce{H2} combination, i.e., \ce{AsH3}/\ce{H2} for InAs nanostructures, and then to room temperature in \ce{N2}.

{\bf Characterisation:} The dimensions, facet determination and morphology of the nanostructures were obtained using either a FEI Verios 460L or a FEI Helios 600 NanoLab field-emission scanning electron microscope with a through lens detector at accelerating voltage between $2$ and $10$~kV and beam current between $50$~pA and $20$~nA. SEM images were recorded at angles of $0^{\circ}$ (top-view), $20^{\circ}$, $30^{\circ}$ and $45^{\circ}$ to normal.

{\bf Nanofin transfer and device fabrication:} The device substrates are $300~\mu$m $2"$ (100) Si wafers doped $n$-type to $0.001-0.005~\Omega$cm. On the front-side, we grow $100$~nm of thermal \ce{SiO2} and then deposit $10$~nm of \ce{HfO2} at $150^{\circ}$C in a Cambridge Nanotech Savannah 100 Atomic Layer Deposition (ALD) system. The \ce{HfO2} layer is not required but included as an etch-stop layer for cases where an oxide-etch is needed in later processing.\cite{StormNL12, GluschkeNanotech19} We protect the front-side with hard-baked photoresist, etch the back-side oxide to completion in buffered \ce{HF}, then deposit $5$~nm Ti and $100$~nm Au by vacuum thermal evaporation to obtain low-resistance contact to the doped substrate, which we use as a global back-gate. After stripping the hard-baked photoresist in hot acetone, we deposited Ti/Au bond-pads, interconnects and alignment markers by one round of photolithography and one round of EBL. This gave $3.5 \times 5.5$~mm chips each with $24$ adjacent device fields ($100 \times 100~\mu$m), each with four contacts in the corners. Corner contacts in adjacent fields are common, such that for a device with $4$ contacts we need $1$ field, $6$ contacts needs $2$ fields, and so on. Device substrates are cleaved to individual chips and thoroughly cleaned by ultrasonication in acetone and 2-propanol prior to use. Mechanical transfer was performed with a micromanipulator system consisting of a high magnification optical microscope (Leica), precision stage (Zaber) and piezo-controlled robot arm (Eppendorf) driving an ultrasharp needle (American Probe Technologies, $0.1~\mu$m radius), combined with some significant practiced skill and patience. The locations of the transferred nanofins relative to the alignment markers are recorded by darkfield microscopy, and used to design appropriate contact and local-gate structures. The device substrate is spin-coated with $950$k-A5 PMMA EBL resist at $5000$~rpm for $60$~s followed by a bake at $180^{\circ}$C for $5$~min on a hot-plate. EBL was performed using a Raith 150-two EBL system (different from templates) with $20$~kV beam energy, $20~\mu$m aperture and $\sim300~\mu$C/cm$^{2}$ typical dose. Development was performed in $1:3$ methylisobutylketone:2-propanol solution for $60$~s for both contacts and local-gates. For the contacts, we perform \ce{(NH4)2S}$_{x}$ passivation at $40^{\circ}$C for $2$~min immediately prior to vacuum evaporation of approximately $5$~nm Ni and $135$~nm Au and liftoff in acetone. The local-gates require two EBL steps: one for the gate-insulator and one for the gate metal. The gate-insulator is $12$~nm of \ce{HfO2} deposited at $100^{\circ}$C by ALD followed by liftoff. The gate metal is approximately $5$~nm Ti and $135$~nm Au by vacuum evaporation followed by liftoff. The completed devices are electrically tested on a probe station, with those viable for further study packaged in LCC20 packages (Spectrum) and bonded with Al wire.

{\bf Electrical measurements:} Electrical measurements were performed with devices mounted on an Oxford Instruments Heliox VL $^{3}$He system loaded into a liquid helium dewar (Wessington CH-120). This system has a small $2$~T superconducting solenoid integrated into the sample-space vacuum can. Temperatures over the range $280$~mK to $30$~K are readily achieved with good control. Data was obtained using standard a.c. lock-in techniques using SR-830 lock-ins for demodulation and $I$-to-$V$ conversion. Channel bias and current were both continuously monitored in addition to other potentials, e.g., Hall, during measurements.
\\
\\
{\bf Supporting Information:} The Supporting Information is available free of charge on the ACS Publications website at http://pubs.acs.org. Additional information including growth characterization, fabrication details and additional electrical data.
\\
\\
{\bf Acknowledgment:} We thank D.J.~Carrad, S.~Upadhyay, J.~Nyg{\aa}rd and N.~Demarina for helpful discussions. This work was funded by the Australian Research Council (ARC) and the University of New South Wales. This work was performed in part using the NSW and ACT nodes of the Australian National Fabrication Facility (ANFF).
\\

\end{document}